\documentclass[aps,prd,showkeys,showpacs,twocolumn,
epsf,preprintnumbers,nofootinbib,superscriptaddress]{revtex4}
\usepackage[dvips]{graphicx}
\usepackage{bm,latexsym,amsmath,amssymb,amsfonts,color}
\usepackage{graphicx}
\usepackage{dcolumn}
\usepackage{bm,morefloats,color}
\usepackage[
      colorlinks=true,
      linkcolor=blue,
      urlcolor=blue,
      filecolor=blue,
      citecolor=red,
      pdfstartview=FitV,
      pdftitle={},
      pdfauthor={},
      pdfsubject={},
      pdfkeywords={},
      pdfpagemode=None,
      bookmarksopen=true
]{hyperref}
\usepackage{amssymb,bm}

\newcommand\beq{\begin{eqnarray}}
\newcommand\eeq{\end{eqnarray}}
\newcommand{\bea}{\begin{eqnarray}}
\newcommand{\eea}{\end{eqnarray}}

\def\X5sp{{\rm X}_5}
\def\Y3sp{{\rm Y}_3}
\def\Z3sp{{\rm Z}_3}

\begin{document}

\title{Thermodynamics of extremal rotating thin shells in an extremal
  BTZ spacetime and the extremal black hole entropy}
\author{Jos\'e P. S. Lemos}
\email{joselemos@ist.utl.pt}
\affiliation{Centro Multidisciplinar de Astrof\'isica - CENTRA,
Departamento de F\'{\i}sica, Instituto Superior T\'ecnico - IST,
Universidade de Lisboa - UL, Avenida Rovisco Pais 1, 1049-001 Lisboa,
Portugal}
\author{Masato Minamitsuji}
\email{masato.minamitsuji@ist.utl.pt}
\affiliation{Centro Multidisciplinar de Astrof\'isica - CENTRA,
Departamento de F\'{\i}sica, Instituto Superior T\'ecnico - IST,
Universidade de Lisboa - UL, Avenida Rovisco Pais 1, 1049-001 Lisboa,
Portugal}
\author{Oleg B. Zaslavskii}
\email{zaslav@ukr.net}
\affiliation{Department of Physics and Technology, Kharkov V. N. Karazin National
University, 4 Svoboda Square, Kharkov 61022, Ukraine, and Institute of
Mathematics and Mechanics, Kazan Federal University, 18 Kremlyovskaya St.,
Kazan 420008, Russia}

\begin{abstract}
In a (2+1)-dimensional spacetime with a negative cosmological
constant, the thermodynamics and the entropy of an extremal rotating
thin shell, i.e., an extremal rotating ring, are investigated. The
outer and inner regions with respect to the shell are taken to be the
Ba\~{n}ados-Teitelbom-Zanelli (BTZ) spacetime and the vacuum ground
state anti-de Sitter (AdS) spacetime, respectively. By applying the
first law of thermodynamics to the extremal thin shell, one shows that
the entropy of the shell is an arbitrary well-behaved function of the
gravitational area $A_+$ alone, $S=S(A_+)$.  When the thin shell
approaches its own gravitational radius $r_+$ and turns into an
extremal rotating BTZ black hole, it is found that the entropy of the
spacetime remains such a function of $A_+$, both when the local
temperature of the shell at the gravitational radius is zero and
nonzero. It is thus vindicated by this analysis, that extremal black
holes, here extremal BTZ black holes, have different properties from
the corresponding nonextremal black holes, which have a definite
entropy, the Bekenstein-Hawking entropy $S(A_+)= \frac{A_+}{4G}$,
where $G$ is the gravitational constant.  It is argued that for
extremal black holes, in particular for extremal BTZ black holes, one
should set $0\leq S(A_+)\leq \frac{A_+}{4G}$, i.e., the extremal black
hole entropy has values in between zero and the maximum
Bekenstein-Hawking entropy $\frac{A_+}{4G}$.  Thus, rather than having
just two entropies for extremal black holes, as previous results
have debated, namely, 0 and $\frac{A_+}{4G}$, it is shown here that
extremal black holes, in particular  extremal BTZ black holes,
may have a continuous range of
entropies, limited by precisely those two entropies.  Surely, the
entropy that a particular extremal black hole picks must depend on
past processes, notably on how it was formed.  A
remarkable relation between the third law of thermodynamics and the
impossibility for a massive body to reach the velocity of light
is also found.  In
addition, in the procedure, it becomes clear that there are two
distinct angular velocities for the shell, the mechanical and
thermodynamic angular velocities. We comment on the relationship
between these two velocities. In passing, we clarify, for a static
spacetime with a thermal shell, the meaning of the Tolman temperature
formula at a generic radius and at the shell.
\end{abstract}

\pacs{
04.70.Dy, 04.40.Nr. 
}
\keywords{Quantum aspects of black holes, thermodynamics, three-dimensional
black holes, spacetimes with fluids}
\date{\today}
\maketitle


\section{Introduction}

The Ba\~{n}ados-Teitelboim-Zanelli (BTZ) spacetime~\cite{btz}
is the spacetime of a (2+1)-dimensional rotating black hole
in a negative cosmological constant $\Lambda$  background,
being thus an asymptotically anti-de Sitter (AdS)
spacetime with length scale $\ell=1/\sqrt{ -\Lambda}$. 
It obeys a no hair theorem \cite{bss}, i.e., 
the black hole is characterized by 
its gravitational radius $r_+$ and its Cauchy radius
$r_-$  or, equivalently, by its
mass $m$ and its
angular momentum $\cal J$. It is thus a simple (2+1)-dimensional
spacetime, and as such 
it provides a way to test and check many different 
properties of the Kerr black holes in the usual
(3+1)-dimensional spacetime.

An important property of black holes is their entropy,
through it one can 
grasp the microscopic
intrinsic elements of a spacetime. 
For extremal BTZ black holes, i.e., black holes 
for which the gravitational radius  is equal to
the Cauchy radius,
$r_+=r_-$,  or the angular momentum is equal to the
mass, ${\cal J}=m\ell$, 
it was found
through topological arguments in the Euclidean sector
 that 
the entropy $S$ is~\cite{ebh2}
\begin{equation}
S=0\,.
\label{ent1ebh}
\end{equation}
Other studies in string theory \cite{birmsacsen}
suggested that the extremal BTZ black hole 
entropy is the Bekenstein-Hawking entropy, namely, 
\begin{equation}
S=\frac{A_+}{4G}\,,
\label{ent1bh}
\end{equation}
where 
$A_{+}=2\pi
r_{+}$ is the event horizon area,
actually here a perimeter,
$r_{+}$ is the gravitational radius,
$G$ is the three-dimensional gravitational 
constant, and we use units such that the
velocity of the light, the Planck constant,
and the Boltzmann constant are set to one.
Thus, the black hole entropy for extremal 
BTZ black holes is not a settled issue.
On the other hand, for nonextremal BTZ black holes, i.e., black holes 
for which the angular momentum is less than the mass, ${\cal J}<m\ell$, or $r_+>r_-$,
the entropy $S$ is
precisely and uniquely given by the 
Bekenstein-Hawking entropy of Eq.~(\ref{ent1bh}).
For further studies on the thermodynamics and entropy 
of BTZ black holes
see 
\cite{thermo_btz1,thermo_btz2,micro_btz2,carlip,micro_btz1,thermo_btz3,ant}.

As the concept of entropy is originally 
based on quantum properties of matter, it
would be useful to study whether the black hole thermodynamics could
emerge from thermodynamics of collapsing matter, when we compress
matter within its own gravitational radius.
So, in order to 
understand better the
physics at the 
horizon, 
a promising setting is a thin shell, i.e., a thin ring, in 
a (2+1)-dimensional spacetime that is compressed
quasistatically to its own gravitational
radius. Outside the 
ring, the spacetime has the BTZ form, inside 
it, the space time is the ground state of the
BTZ spacetime, i.e., a zero mass locally 
AdS spacetime.
One can calculate the entropy of this matter 
ring system for any ring radius $R$, 
in particular, when the ring is compressed to its  
gravitational radius $r_+$, $R=r_+$.
This has been done in the  nonrotating
BTZ case \cite{quintalemosbtzshell} 
and in the rotating nonextremal BTZ case 
\cite{btzshell} where the entropic 
properties of the ring at the gravitational radius
were deduced. In particular, it was found 
that the entropy of the ring is the Bekenstein-Hawking
entropy given in Eq.~(\ref{ent1bh}),
provided that the shell's temperature coincides with the Hawking 
temperature of the corresponding black hole.
Still lacking is the study of the thermodynamics and
the entropy 
of the extremal BTZ ring case, that
might emulate the direct calculation
of the entropy of an extremal BTZ black hole.
For other studies related to 
the properties of matter systems, in particular, 
rotational properties in 
BTZ backgrounds
see \cite{collapse1,collapse2,mop,kimporrati,energycond}.

The fact that shells 
reflect black hole properties was found
in (3+1) Reissner-Nordstr\"om
asymptotically flat spacetimes 
with an electric shell
where the entropy properties of
zero charge, i.e., Schwarzschild
\cite{martin},
nonextremal \cite{charged},
and extremal \cite{extremalshell,lqzn},
black holes were reproduced, making these shells 
a very useful setting.
Related to it 
there were the studies of 
the entropy for  quasiblack holes 
\cite{quasi_bh1,quasi_bh2}
and of quasistatic collapse of matter \cite{pretisrvol}.
In the nonextremal case these studies found
that at the gravitational radius of the shell the spacetime,
and thus the corresponding black hole,
has the Bekenstein-Hawking entropy of Eq.~(\ref{ent1bh}),
where in the (3+1)-dimensional case
$A_+=4\pi\,r_+^2$.
On the other hand, for extremal shells 
the entropy at the gravitational radius
and thus the entropy of the corresponding
extremal black hole, 
can be any well-behaved function
of the gravitational radius $r_+$, or
since $A_+=4\pi\,r_+^2$, 
the entropy can be any well-behaved function
of the gravitational area $A_+$, 
$S=S(A_+)$. So among many other values, 
it can be zero as in Eq.~(\ref{ent1ebh})
or
$\frac{A_+}{4G}$ as in Eq.~(\ref{ent1bh}).

The ambiguity in the entropy of extremal black holes, that can be
either zero or the Bekenstein-Hawking entropy, was indeed found first
in (3+1)-dimensional black holes.  For extremal (3+1) black holes, it
was found in one approach based on the horizon topology \cite{ebh1}
(see also~\cite{ebh2}) that the entropy is zero $S=0$, see
Eq.~(\ref{ent1ebh}).  The other approach, based on string theory
calculations, yields that the entropy of extremal black holes is given
by the Bekenstein-Hawking entropy,
Eq.~(\ref{ent1bh})~\cite{string1,string2}, see also
\cite{ebh3,ebh4,ebh5,ebh6,ebh7,ghoshmitra,string3,string4,string5,cano1}.
On the other hand, for (3+1)-dimensional nonextremal black holes, the
entropy is the original unambiguous Bekenstein-Hawking entropy,
$S=\frac{A_+}{4G}$ of Eq.~(\ref{ent1bh}) \cite{bek1,bch,haw}.

Here, we pursue further the problem of the entropy of an 
extremal BTZ black hole by using a shell, an extremal 
rotating thin shell. 
This is important in order to gain new insights
into the entropy and other physical relevant quantities
from spacetimes that possess rotation and 
angular momentum.

The paper is organized as follows. 
In Sec.~\ref{sec2}, we discuss the mechanics of an extremal rotating
thin shell in (2+1) dimensions with a cosmological
constant. The exterior spacetime to the shell is the BTZ
spacetime.
In Sec.~\ref{sec3}, we study the first 
law of thermodynamics applied for such a thin shell,
derive the thermodynamic entropy of the thin shell,
and show that the entropy
is a function of the gravitational area
$A_+$ only, $S=S(A_+)$. We also analyze the equation of
state for the temperature and for the angular
velocity of the shell.
In Sec.~\ref{sec4}, we consider the extremal shell with zero local
temperature and take the limit to its gravitational radius,
obtaining thus the properties of the
corresponding extremal black hole.
In Sec.~\ref{sec5}, we consider the extremal shell with some
nonzero
constant local temperature and take again
the limit to its gravitational radius,
obtaining also the properties of this
extremal black hole.
In Sec.~\ref{app1}, we discuss the nontrivial relation between
mechanical and thermal, angular and linear, velocities and compare the
nonextremal and extremal cases.
In Sec.~\ref{sec7}, we give some 
concluding remarks. In the Appendix \ref{tolmangeneral},
we 
clarify the meaning of the
Tolman temperature formula at a generic radius and at the shell
for a spacetime with a thermal shell.

\section{Thin shells in a (2+1)-dimensional extremal BTZ
spacetime}
\label{sec2}

\subsection{Outer and inner spacetimes}

We consider general relativity with
a cosmological constant $\Lambda$
in a (2+1)-dimensional spacetime.
We also assume that $\Lambda<0$, so that the spacetime is asymptotically
AdS, with curvature length scale $\ell=1/\sqrt{ -\Lambda}$.
Throughout this paper, we work in units where
the velocity of light, the Planck constant, and
the Boltzmann constant are set to unity.~$G$
denotes the gravitational constant in (2+1) dimensions.

We introduce a timelike shell, i.e., a ring, in the (2+1)-dimensional
spacetime, with radius $R$, which divides the spacetime into the outer
and inner regions labelled by $(o)$ and $(i)$, respectively
\cite{quintalemosbtzshell,btzshell} (see also \cite{energycond}).
We assume that the
spacetime is vacuum everywhere off the shell. Outside the shell
($r>R$), the spacetime is described by the extremal rotating BTZ
solution, while inside the shell ($r<R$), the spacetime is
the ground state of the BTZ solution and is locally AdS.  One
can express the line element for the inner and outer regions by
\begin{eqnarray}  \label{collect}
ds_{(I)}^2 &=&-f_{(I)} (r)\,dt_I^2 + g_{(I)}(r)\,dr^2  \notag \\
&+&r^2 \big(d\phi-\omega_{(I)} (r)dt_{(I)}\big)^2,
\end{eqnarray}
where $t$ is the time coordinate,
$(r,\phi)$ are the radial and azimuthal coordinates,
$I=o/i$ refers to the outer or inner region in relation to the shell,
respectively,
and the functions $f_{(I)}$, $g_{(I)}$ and $\omega_{(I)}$ read 
\begin{eqnarray}
f_{(o)}(r)=&&
\left(\frac{r} {\ell}\right)^2
\Big(1-\frac{r_+^2}{r^2}\Big)^2\,,\quad
g_{(o)}(r)=\frac{1}{f_{(o)}(r)}, \nonumber\\
&& \omega_{(o)}(r)= 
\frac{r_+^2}{\ell r^2},  \label{outq}
\\
f_{(i)}(r)=&&
\left(\frac{r} {\ell}\right)^2,\quad\;\;\;\; 
\quad g_{(i)}(r)=\frac{1}{f_{(i)}(r)},\;\; \nonumber\\
&& \omega_{(i)}(r)=0. \label{inq}
\end{eqnarray}
The subscript $(I)$ in the time coordinate $t_{(I)}$ indicates that in
general the time coordinate of the outer region $t_{(o)}$ differs from
that of the inner region $t_{(i)}$.
The radius $r_{+}$ is
the gravitational radius  of
the spacetime. In the extremal case, the case we consider here,
the gravitational radius $r_+$ is equal to the 
 Cauchy radius $r_-$, $r_+=r_-$, and we stick to the usual notation
 $r_+$ for such a radius. The gravitational radius $r_{+}$
becomes the horizon radius if the
solution is an extremal black hole or an object on the verge of
becoming an extremal black hole. 
In the extremal case 
the
radius $r_{+}$ is given by 
\begin{equation}
\label{ml}
r_{+}^{2}=4G\ell ^{2}m\,,
\end{equation}
where $m$ is the asymptotic Arnowitt-Deser-Misner (ADM)
mass. The radius $r_{+}$
can also be written as $r_{+}^{2}=4G\ell \mathcal{J}$
upon using that the spacetime
angular momentum $\mathcal{J}$ and  $m$ are related 
in the extremal case by
\begin{equation}
\label{mjl}
\mathcal{J}=m\,\ell\,.
\end{equation}
We assume $m>0$.
A gravitational or horizon radius $r_+$ 
corresponds to a gravitational or horizon area $A_+$, 
here a perimeter, given by
\begin{equation}
\label{a+r+}
A_+=2\pi r_{+}\,.
\end{equation}
The inner region $(i)$ corresponds to the
ground state vacuum solution, i.e., 
$m=0$ and $\mathcal{J}=0$.
In the junction between the outer extremal BTZ spacetime and 
the inner vacuum AdS spacetime, at the radius $R$,
there is a stationary thin shell.
We assume that the shell's character is always timelike and the shell is
located outside the event horizon, 
\begin{equation}
\label{minim}
R\geq r_{+}\,.
\end{equation}
Therefore, the outer extremal region $(o)$ does not contain an event
horizon $r=r_{+}$, except in the case $R= r_{+}$.  One can also define
the area $A$, here a perimeter, of the shell as
\begin{equation}
\label{AR}
A=2\pi R\,,
\end{equation}
so that Eq.~(\ref{minim}) is written as
\begin{equation}
\label{minima}
A\geq A_{+}\,.
\end{equation}

An important quantity is the redshift function $k$
at some coordinate outer radius $r$,
$k\equiv \sqrt{f_{(o)}(r)}$
which in our case is 
\begin{eqnarray}\label{reds0}
k\big(r_+,r\big) =\frac{r}{\ell}\Big(1-\frac{r_+^2}{r^2}\Big).
\end{eqnarray}
This function gets a value equal to one at the coordinate $r=r_0$
given by
\begin{eqnarray}  
\label{rastr}
r_0 = \frac{\ell}{2} +\sqrt{\Big(\frac{\ell}{2}\Big)^2+r_+^2},
\end{eqnarray}
where $k(r_+,r_0)=1$. 
It is also of interest to display the redshift function 
at the position of the shell, 
$k\equiv \sqrt{f_{(o)}(R)}$,
which 
in our case is 
\begin{eqnarray}\label{reds}
k\big(r_+,R\big) =\frac{R}{\ell}\Big(1-\frac{r_+^2}{R^2}\Big).
\end{eqnarray}
This function gets a value equal to one when  
the shell is  at the position
\begin{eqnarray}  
\label{rast}
R_0 = \frac{\ell}{2} +\sqrt{\Big(\frac{\ell}{2}\Big)^2+r_+^2},
\end{eqnarray}
where $k(r_+,R_0)=1$, and $R_0$ is always greater than both $r_+$ and 
$\ell$.

\subsection{Spacetime at the junction: Properties
of the extremal shell}

\subsubsection{Metric and rotation of the extremal shell}

The shell dynamics and its matter content 
are  determined by the Israel junction conditions. 
The first junction condition ensures the uniqueness of
the induced
geometry on the shell, at $R$, $h_{ab}=h^{(o)}_{ab}=h^{(i)}_{ab}$,
where $a,b=t,\phi$. The second
junction condition determines the energy-momentum tensor of matter on the
thin shell, $S_{ab}$, that compensates the jump of the extrinsic curvature
tensor across the shell.

As the outer spacetime is rotating while the inner spacetime is static, in
order to match these two regions, the shell at $r=R$ must corotate with the
outer BTZ region. We introduce a coordinate system corotating
with the shell by adopting a new angular coordinate $d\psi$ such that 
\begin{equation}
d\psi= d\phi-\omega_{(I)} (R)dt_{(I)}\,.
\label{ct}
\end{equation}
The line element given in Eq.~\eqref{collect} is
then written as 
\begin{eqnarray}
ds_{(I)}^2 &=& -f_{(I)} (r)dt_{(I)}^2 +g_{(I)}(r) dr^2  \notag \\
&+& r^2\big(d\psi -{o}_{(I)} (r)dt_{(I)}\big)^2,
\end{eqnarray}
where we have introduced 
\begin{equation}
o_{(I)}(r)= \omega_{(I)} (r)-\omega_{(I)} (R)\,,
\label{o}
\end{equation}
so that at the position of the shell ${o}_{(I)}(R)=0$ and the 
line element is diagonal. Also, 
from Eqs.~(\ref{outq}) and (\ref{inq}),
\begin{equation}
\omega_{(o)} (R)=\frac{r_+^2}{ 
\ell R^2}\,,
\label{shellrotout}
\end{equation}
and 
\begin{equation}
\omega_{(i)} (R)=0\,,
\label{shellrotin}
\end{equation}
respectively.
The
induced line element on the shell at $R$ uniquely determined by~the first
junction condition is given by
\begin{equation}  \label{ind}
ds_R^2  =-d\tau^2+R^2d\psi^2, \quad R=R(\tau)\,.
\end{equation}
The proper time on the shell $\tau$ is
defined by 
\begin{eqnarray}  \label{ind2}
d\tau &=\sqrt{ f_{(o)} (R)dt_{(o)}^2 -g_{(o)}(R) dR^2} \notag \\
&=\sqrt{
f_{(i)} (R)dt_{(i)}^2 -g_{(i)}(R) dR^2}\,.
\end{eqnarray}
Since we are interested in a
quasistatic process  we always assume that $\frac{dR}{d\tau}=0$ and 
$\frac{d^2R}{d\tau^2}=0$.

\subsubsection{Energy-momentum tensor of the extremal shell}

The rotating thin shell is supported by an imperfect fluid with an
energy-momentum tensor $S^{a}{}_{b}$, such that the
nonzero components are 
$S^{\tau}{}_{\tau}= -\sigma$, $S^{\psi}{}_{\psi}=p$, and
$S^{\tau}{}_{\psi}=j$, where $\sigma$, $p$, and $j$, represent the
energy density of the shell, the pressure in the shell, and the
 angular momentum flux density of the shell, respectively.
The second
junction condition gives 
\begin{eqnarray}
\sigma &=& \frac{r_+^2}{8\pi G \ell R^2}\,,  \label{ss1}\\
j&=&\frac{r_+^2}{8\pi G \ell R}\,,  \label{j1}\\
p &=&\frac{r_+^2}{8\pi G \ell R^2}\,.  \label{p1} 
\end{eqnarray}
Thus, $\sigma=p=j/R$. The extremal rotating shell obeys both the weak
and dominant energy conditions~\cite{energycond}.

Defining the local proper mass and angular momentum of the shell by $M=2\pi
R\sigma$ and $J=2\pi R j$, respectively, and using Eqs.~(\ref{ss1}) and (\ref
{j1}), we obtain 
\begin{eqnarray}
M &=& 2\pi R\sigma =\frac{r_+^2}{4G\ell R},  \label{Mmass1} \\
J &=& 2\pi R j = \frac{ r_+^2}{4G \ell}\,.  \label{JJ1}
\end{eqnarray}

\subsubsection{Relation between global extremal 
spacetime quantities and local 
extremal shell quantities}

The spacetime quantities $m$ and $\cal J$ are
related to the shell quantities $M$ and $J$.
From Eqs.~\eqref{ml} and \eqref{Mmass1},
one finds that 
the local proper mass of the shell $M$ is related to the ADM mass $m$ by 
\begin{eqnarray}  \label{ex1}
m=\frac{MR}{\ell}.
\end{eqnarray}
From Eq.~\eqref{JJ1}, one sees
that the angular momentum of the shell $J$ is independent of the position
of the shell $R$, and from Eqs.~(\ref{ml}) and(\ref{mjl}),
we see that it is identical to
the angular momentum  of the outer BTZ spacetime, 
\begin{eqnarray}
\mathcal{J} =J\,.
\end{eqnarray}

We would like to emphasize that in our case the inner region is a
$(2+1)$-dimensional spacetime locally AdS and it has  zero
ADM mass and zero angular momentum. In the more complex case that the
region inside the shell contains instead a BTZ black hole, then the
total ADM mass and angular momentum of the outer spacetime defined at
infinity would include in addition the ADM mass and angular momentum
of the interior black hole.

\section{First law of thermodynamics, entropy
of an extremal rotating thin shell in a BTZ spacetime, and
the equations of state}
\label{sec3}

\subsection{First law of thermodynamics of an
extremal rotating thin shell in a BTZ spacetime}

Now, we turn to the thermodynamics of the thin shell. Following \cite
{btzshell}, we assume that the rotating
shell has a thermodynamic angular velocity
$\Omega$ and is in thermal equilibrium, with 
local temperature $T$ and entropy $S$. The entropy $S$ of a system can be
expressed as a function of the state independent variables. One can take as
state independent thermodynamic variables for the thin shell the proper
mass $M$, the area of the shell $A$, and the angular momentum
of the shell $J$. 
Thus, the entropy of the shell
is a function of these quantities 
through the first law of thermodynamics which reads 
\begin{eqnarray}  
\label{1st00}
TdS=dM + p \, d A-\Omega\, dJ\,.
\end{eqnarray}
To obtain the entropy 
$S$, in general we have to specify the three equations of state $T(M,A,J)$, $
p(M,A,J)$, and $\Omega(M,A,J)$, namely, the temperature, pressure, and
angular velocity equations of state, respectively. 
To proceed in this direction, define the inverse local temperature
of the shell as 
\begin{equation}  \label{it}
\beta= \frac1T\,.
\end{equation}
Note further that in a
(2+1)-dimensional spacetime the area $A$ of the shell,
actually a perimeter in
common usage, is given by Eq.~\eqref{AR} 
so that it is mathematically equivalent to the position $R$
except for the trivial factor $2\pi$,  i.e., we can 
make use of the variable $R$ instead of $A$.
Using Eq.~\eqref{it},  the first law of thermodynamics (\ref{1st00})
now reads in the ($M,R,J$) variables
\begin{eqnarray}  \label{1st}
dS=\beta\, dM+2\pi \beta p \, dR - \beta\Omega\, dJ.
\end{eqnarray}
We need to give the equations of state for
$\beta(M,R,J)$, $
p(M,R,J)$, and $\Omega(M,R,J)$.

\subsection{Entropy of an extremal rotating thin shell in a BTZ spacetime}

We can make progress using first, for the time
being, the equation of state for the pressure $p$.
Through the junction condition, i.e., through
the spacetime mechanics, $p$ is indeed fixed by 
Eq.~\eqref{p1}.
Changing to the variables 
$M$ and $R$, and
using Eqs.~\eqref{p1} and \eqref{Mmass1}, valid for an extremal shell,
one finds that the equation of state for the pressure $p$
can be written as 
\begin{equation}
p(M,R)=\frac{M}{2\pi R}\,.
\label{pMR}
\end{equation}
Also, one can still take advantage
of Eqs.~\eqref{Mmass1} and \eqref{JJ1}. Through these
equations, we obtain
\begin{equation}
J=MR\,.
\label{ex3}
\end{equation}
Equation~\eqref{ex3} gives that $J$, $M$, and $R$ are not
independent, with $dJ=RdM+MdR$.
Putting Eqs.~\eqref{pMR} and \eqref{ex3}
into the first law Eq.~\eqref{1st},
we obtain the differential of entropy 
$dS=\frac{\beta }{R}\big(1-\Omega R\big)d(MR)$. It is then useful
to define the the thermodynamic rotational velocity of the shell $V$
by 
\begin{equation}
V=\Omega R \,, \quad\quad\quad\quad 0\leq V\leq 1\,,
\end{equation}
where the  constraint $V\leq 1$ ensures 
that the maximum velocity is the velocity of light.
Then, the entropy of the extremal rotating shell obeys
\begin{equation} 
\label{ent_ext0}
dS=\frac{\beta }{R}\big(1-V\big)d(MR)\,.
\end{equation}

Clearly, from Eq.~\eqref{Mmass1}, i.e., $MR=\frac{
r_{+}^{2}}{4G\ell }$, it is natural to pass from the 
variables $MR$ to the variable $r_+$. In this 
variable $r_+$, the differential of
entropy Eq.~\eqref{ent_ext0} can further be reduced to 
\begin{equation}  \label{ent_ext}
dS=\frac{r_{+}}{2G\ell }\frac{\beta }{R}\big(1-V\big)dr_{+}.
\end{equation}
It is manifest that Eq.~\eqref{ent_ext} has to be written as
\begin{equation}  \label{ent_ext2}
dS=s(r_{+})dr_{+}, 
\end{equation}
where 
\begin{equation}
s(r_{+})=\frac{r_{+}}{2G\ell }\frac{\beta }{R}\big(1-V\big),
\label{sp}
\end{equation}
is actually the integrability condition for
Eq.~\eqref{ent_ext} and
where $\beta$ and $V$ are arbitrary functions of $(r_{+},R)$, but
$s(r_{+})$ is an arbitrary function of $r_{+}$ alone.
Thus, the relation~\eqref{ent_ext2} indicates that the entropy $S$
of the shell is a
function of the gravitational radius of the shell $r_{+}$ only, 
\begin{equation}  \label{entrp}
S=S(r_{+}),\quad\quad R>r_{+}\,,
\end{equation}
where we have set a constant of integration to zero.
Since one can trade $r_+$ for $A_+$ trivially through
Eq.~\eqref{a+r+}, we write Eq.~\eqref{entrp} in the more
visual form
\begin{equation}  \label{entrparea}
S=S(A_{+}),\quad\quad A>A_{+}.
\end{equation}
Depending on the choice of $s(r_{+})$ in Eq.~\eqref{sp},
the entropy $S(r_{+})$, or equivalently $S(A_{+})$,
of the rotating shell in the BTZ spacetime 
can take a wide range of values. 

Now, in the variables $(r_+,R)$ (or if one prefers,
$(A_+,A)$), which are the 
natural variables for calculation in this problem, one
has two remaining functions of state, 
$\beta (r_{+},R)$ and $V(r_{+},R)$. These two
functions of state, $\beta (r_{+},R)$ and $V(r_{+},R)$,
are arbitrary as long as they obey the 
thermodynamic constraint \eqref{sp}. The extremal
rotating shell is in this sense quite special.

\subsection{Equations of state for the inverse temperature $\beta$
and for the rotational velocity $V$}

\subsubsection{Inverse temperature $\beta$ equation of state }

Now, we prepare a given extremal shell at a generic radius $R$.
The unique
integrability condition \eqref{sp} is quite general and does not
impose per se a restriction on the temperature distribution. 
However, thermodynamic systems have to obey the Tolman 
formula for the  temperature or, equivalently, 
for the inverse local temperature.

The Tolman temperature formula,
by its very meaning, states that
the coordinate dependence of the temperature
obeys some restriction.
The Tolman temperature formula for spherical
systems of the type we are using here is
$T(r) = T_0/k(r)$
where the dependence on $r$, 
the local radial coordinate, exists
only in the redshift factor $k$, see Appendix \ref{tolmangeneral} for a 
thorough discussion on the Tolman formula for shells.
The quantity $T_0$ may depend on the parameters of the system. 
In our case, the gravitational radius $r_+$ and 
the radius of the shell $R$ are examples of such parameters,
such that $T_0=T_0(r_+,R)$.
For a given
shell's position, fixed $R$, we can consider the Tolman formula in the
whole outer space.  For some other $R$, the configuration changes, and for
that new configuration we can consider again the Tolman 
formula in the new
whole outer space.  More precisely,
suppose that we have a shell at position $R$ 
with intrinsic 
gravitational radius $r_+$ at some given
temperature. The  local
temperature of the spacetime at some specific radius 
$r$ is
$T$, say.
The
Tolman formula relates the local temperature $T$ to a constant
temperature parameter $T_0$. In the case under discussion, $T_0$ has
the meaning of the local temperature at $r=r_0$, where $k(r_0,r_+)=1$
(see Eq.~(\ref{rastr})).
In brief, the Tolman formula states that $T$ at $r$, $T(r)$, is a 
function of a temperature $T_0$, at the radius $r_0$
where $k(r_0,r_+)=1$. $T_0$
is itself a function of the 
characteristics of the system, $r_+$ and $R$ in our case,
\begin{equation}  \label{temp0}
T_0=T_0(r_+,R)\,.
\end{equation}
Thus, dividing by the redshift function at $r$ given in
Eq.~(\ref{reds0}),  
the Tolman temperature
formula in full is
\begin{equation}  \label{temptolman0}
T(r_+,R,r)=\frac{T_0(r_+,R)}{k(r_{+},r)}\,.
\end{equation}
Inverting this equation, and using  Eq.~(\ref{it}), yields
the required Tolman formula for the inverse temperature $\beta$, i.e., 
\begin{equation}  \label{temp_ext00}
\beta(r_+,R,r)=b(r_+,R) k(r_{+},r),
\end{equation}
where
\begin{equation}  \label{invtemptob0}
b(r_+,R)=\frac{1}{T_0(r_+,R)}\,.
\end{equation}
In this way, one interprets $b(r_+,R)$
as the inverse temperature at the radius $r=r_0$ 
for which $k=1$. Conversely, $\beta$ is the 
inverse temperature at $r$, blueshifted or redshifted 
with factor $k$ 
from the inverse temperature at the position where 
$k=1$.

Now, on the shell, $r=R$, so the Tolman formula there
is $\beta(r_+,R,r=R)=b(r_+,R) k(r_{+},r=R)$, or
simplifying the notation,
\begin{equation}  \label{temp_ext}
\beta(r_+,R)=b(r_+,R) k(r_{+},R).
\end{equation}
If the shell happens to be at 
$R =R_0$, where 
$R_0= \frac{\ell}{2} +\sqrt{\Big(\frac{\ell}{2}\Big)^2+r_+^2}$,
then $k(r_+,R_0)=1$, see Eq. (\ref{rast}), 
and so there $\beta(r_+,R_0)=b(r_+,R_0)$.

It is important to note that the
formula~\eqref{temp_ext} for 
ab initio extremal shells is more comprehensive,
and so different, than the one found for 
extremal shells formed from taking the limit 
of nonextremal shells \cite{btzshell}.
In \cite{btzshell} it was found 
that for nonextremal shells, with radius $R$
and gravitational and Cauchy radii
$r_+$ and $r_-$, the following Tolman equation 
at the shell's radius $R$, found
from the integrability conditions, holds, i.e., 
$\beta(r_+,r_-,R)=b(r_+,r_-) k(r_{+},r_-,R)$,
where $k\big(r_+,r_-,R\big)
=\frac{R}{\ell}
\sqrt{
\Big(1-\frac{r_+^2}{R^2}\Big)
\Big(1-\frac{r_-^2}{R^2}\Big)}$ is the 
redshift function in 
the nonextremal case. 
Taking then, from the nonextremal shell,
the extremal shell limit $r_+\to r_-$ and
noting that $k\big(r_+,r_-,R\big)=k\big(r_+,R\big)$
in this limit where $k=k\big(r_+,R\big)$ is 
given in Eq.~\eqref{reds}, one finds
$\beta(r_+,R)=b(r_+) k(r_{+},R)$ \cite{btzshell}.
Note the difference: the limit of a nonextremal shell 
to an extremal shell gives $b=b(r_+)$, and $b$ is a function
of $r_+$ alone \cite{btzshell}. On the other hand, when one 
has an extremal shell ab initio, one finds
$b=b(r_+,R)$, see Eq.~\eqref{temp_ext}, i.e., in this more comprehensive
case, 
$b$ is a function
not only of $r_+$ but also of $R$.

The difference 
comes of course from the 
different integrability conditions
arising in the nonextremal and extremal cases.
For an ab initio extremal shell, the only 
integrability condition \eqref{sp} is too general 
and gives $b=b(r_+,R)$ as in Eq.~\eqref{temp_ext}. 
For a nonextremal shell, the three integrabilty 
conditions are very restrictive,
and when one takes the extremal limit, the 
memory of this restrictiveness remains, so $b=b(r_+)$.

\subsubsection{Rotational velocity $V$ equation of state}

With the choice for the inverse temperature equation of state
\eqref{temp_ext}, we find
from Eq.~\eqref{sp} that the rotational velocity of the
shell has the functional form
\begin{eqnarray}
\label{omega_ext}
V(r_+,R)=\frac{R }{\ell k(r_+,R)} \Big( g(r_+,R)-\frac{r_+^2}{R^2} \Big),
\end{eqnarray}
where we have defined 
\begin{eqnarray}
\label{cdef}
g(r_+,R)= 1-\frac{2G\ell^2}{r_+ b(r_+,R)}s(r_+).
\end{eqnarray}
For an ab initio extremal shell, 
$g$ has a dependence on $r_+$ as well as on $R$.
For a nonextremal shell 
and taking the limit 
to the extremal shell \cite{btzshell} the corresponding function
depends only on $r_+$. As in the equation of state
for the inverse temperature, this comes from the very
different integrability conditions in each case.
Note that $g$ in Eq.~\eqref{cdef} corresponds to $c$ in 
Eq.~(59) of \cite{btzshell} 
with  $g=c{r_+^2}$,
but now here
$g$ in general has the dependence on $R$ as well as $r_+$
due to the different integrability conditions as 
discussed.

With the definition of $g$ in Eq.~\eqref{cdef},
we have from Eq.~\eqref{sp} the useful formula 
\begin{eqnarray} 
\label{alf}
s(r_+)=\frac{r_+}{2G\ell^2} b(r_+,R) \big(1-g(r_+,R)\big) \,,
\end{eqnarray}
which shows that, although $b$ and $g$ are functions of 
$r_+$ and $R$, their combination is a function
of $r_+$ alone. 
Also, from the definition of $g$ in Eq.~\eqref{cdef},
we have another useful formula,
\begin{eqnarray} 
\label{1-v}
V\big(r_+,R\big) =1-\frac{R}{\ell k(r_+,R)} \big(1-g(r_+,R)\big)\,.
\end{eqnarray}
We see that the velocity $V\rightarrow 1$ when $g\to1$, i.e., 
when $b\to\infty$ according to Eq.~(\ref{cdef}), or
$T_0\rightarrow 0$. 
In this respect, there is a direct remarkable interesting
relation between the unattainability of the absolute zero of temperature
and the impossibility for a material body to reach the
velocity of light.

\subsection{Explicit computation of the entropy of the shell}
\label{subexpl}

For the explicit computation of the entropy $S$ of the shell,
see Eq.~\eqref{entrp} (or Eq.~\eqref{entrparea}), we have to
specify the equations $b(r_+, R)$ and $g(r_+, R)$ which determine the
thermodynamic properties of the shell. In this paper, we do not proceed
in this way but, instead, focus on Eqs. \eqref{temp_ext} and
\eqref{omega_ext} and study the particular cases for which we can
take the limit to the extremal black hole, $R\to r_+$.

\section{Entropy in the extremal BTZ black hole limit: 
Extremal thin shell with zero local temperature $T$
at the gravitational radius
}
\label{sec4}

We will now study 
the extremal 
black hole limit in the sense that we 
take quasistatically the extremal shell 
to its own gravitational radius, $R=r_+$.

In this procedure of going quasistatically to the gravitational radius
$ R=r_{+}$, we have to prepare in advance the shell. In first place, we
put the shell at some radius $R>r_{+}$ and in addition choose the
functions
$\beta $ and $V$, or $b$ and $g$, in an appropriate
manner. After doing this, we take the shell to $R=r_{+}$. In second
place, we stipulate $b$ and so $\beta $.  We know that the Hawking
temperature $T_{H}$ for a BTZ black hole is measured at $r_{0}$, i.e.,
$T_{0}=T_{H}$, see e.g., \cite{thermo_btz2}. For an extremal black
hole, this temperature is zero $T_{H}=0$. We assume that the equality
$T_{0}=T_{H}$ is valid for our shell since otherwise the backreaction of
quantum fields would destroy it when the shell approaches the horizon.
Now, the temperature $T_{0}$ is precisely related to our $b$,
$b(r_{+},R)=1/T_0(r_{+},R)$, see Eq.~(\ref{invtemptob0}).
But since $T_0=T_H=0$, we have, for a
shell at radius $R$, to set $b=\infty $.  Thus, from
Eq.~(\ref{temp_ext00}), $\beta (r_{+},R,r)=\infty $ and in particular
$\beta (r_{+},R,r=R)\equiv \beta (r_{+},R)=\infty $. The temperature
at the shell is zero. We can now change the radius of the shell
quasistatically, and
the same rationale applies, since we always want $T_0=T_{H}=0$,
i.e., $b=\infty$. In 
third place, we find $g$ and $V$. From Eq.~(\ref{alf}), we find that
when $b=\infty $ then $g=1$. Let us suppose that we start with a
configuration in which $g$ is not equal to $1$ exactly and $b$ is
large but finite. Then, $1-g=s\,(2G\ell ^{2}/r_{+})/b$, for some
well-specified $s$.  We are interested in the limit in which $b\rightarrow
\infty $, $g\rightarrow 1$, $V\rightarrow 1$. To trace in more detail
this limit, we can choose $g$ as close to 1 as we want and $T_{0}$ as
small as we like, i.e., $ b$ as large as we like. In the end, keeping
the product fixed in Eq.~(\ref {alf}) (for a given $r_{+}$), we can
take the limit of $g$ to 1 and $T_0$ to zero, i.e., $b$ to infinity. We
see that the shell at $R>r_{+}$ has been prepared with $T_{0}=0$, i.e.,
$b=\infty$, and $g=1$, such that $\frac{r_{+}}{2G\ell ^{2}}b\big(
1-g\big)=s$ and so $\frac{r_{+}}{2G\ell }\frac{\beta
}{R}\big(1-V\big)=s$, with thus $\beta =\infty $ and $V=1$.  Since
$V=1$, the shell rotates with the velocity of light precisely in this
limit. The shell is now correctly prepared. Having made the correct
preparations on the shell, and as long as $R\geq r_{+}$ and imposing
that Eq.~\eqref{alf} is always obeyed for some fixed $s(r_{+})$, we can
send it to its gravitational radius $r_{+}$.

Let us then send the extremal shell to 
its own gravitational radius $R=r_+$. In doing so 
we are taking the extremal black hole limit. 
Since the entropy differential for the shell
depends only on $r_+$ through the
function $s(r_+)$ that is arbitrary, see
Eq.~(\ref{sp}),
we see that the entropy of
the extremal shell in the extremal black hole limit is given by
\begin{equation}  \label{entrextremalbh}
S=S(r_+) \,,\quad R= r_+\,,
\end{equation}
or, in terms of the horizon area if one prefers, 
\begin{equation}  \label{entrextremalbharea}
S=S(A_+) \,,\quad A= A_+\,.
\end{equation}
This is the extremal black hole limit 
of an extremal shell. This type of configuration, 
a matter system at its own gravitational radius, is 
called
a quasiblack hole \cite{quasi_bh1,quasi_bh2}.
Thus the entropy of the extremal black hole can be any well-behaved
function of $r_+$, or $A_+$,
which depends on the constitution of matter that
collapses to form the extremal BTZ black hole. Depending on the choice
of $s(r_+)$, in turn of $ \beta$ and $V$, we can obtain any function of
$r_+$, or $A_+$, for the entropy $S$ of the extremal black hole.  This result is
quite different from the nonextremal case discussed in 
\cite{btzshell}, where the entropy of the shell for which the temperature
coincides with the Hawking temperature can only take the form of the
Bekenstein-Hawking entropy $S(A_{+})=\frac{A_{+}}{4G}$.

Our preceding calculations and discussion were
exact.  Now, we can speculate
on ways to constrain the entropy function $S(A_+)$
for the extremal black hole.
For the nonextremal 
black holes, the entropy is 
$S(A_{+})=\frac{A_{+}}{4G}$. This expression is found
when one takes the shell to its
own gravitational radius and assumes that the shell takes the Hawking
temperature. In this case,
the pressure at the shell blows up, $p\rightarrow
\infty $ \cite{quintalemosbtzshell,btzshell}.
This blowing up of the pressure can be interpreted
as the excitation of all possible degrees of freedom
and the corresponding black hole takes the Bekenstein-Hawking entropy, 
the maximum possible entropy. Taking the extremal limit
from a nonextremal black hole, one finds 
that in this particular limit the extremal black hole
entropy is the Bekenstein-Hawking entropy \cite{quintalemosbtzshell,btzshell}
(see also \cite{quasi_bh1,quasi_bh2}).
Thus, this suggests that 
the maximum entropy that an extremal black hole
can take is the Bekenstein-Hawking entropy. 
Therefore, in this sense,
the range of values for the entropy of an
extremal black hole is
\begin{equation}  \label{Srange0}
0\leq S(r_{+})\leq \frac{\pi r_+}{2G}\,,
\end{equation}
or, in terms of $A_+$,
\begin{equation}  \label{Srange}
0\leq S(A_{+})\leq \frac{A_+}{4G}\,.
\end{equation}

The result \eqref{entrextremalbh}, or equivalently
\eqref{entrextremalbharea},
is a quite similar result to the
case of the extremal charged shell in a (3+1) Reissner-Nordstr\"om
spacetime \cite{extremalshell,lqzn}.  As for the extremal electrically
charged shells \cite{extremalshell,lqzn}, we then constrained the
entropy function $S(r_+)$, or $S(A_+)$,
for the extremal black hole. For the
nonextremal Reissner-Nordstr\"om
black holes, the entropy is given by the
Bekenstein-Hawking formula. In this case, when the shell is
taken to its own gravitational radius, the pressure at the shell
diverges, $p\to \infty$ as $ k^{-1}$ (see (54) of \cite{btzshell}),
and the spacetime is assumed to take the Hawking temperature. Thus all
possible degrees of freedom on the shell are excited and the black
hole formed as the limit of the shell to its gravitational radius
takes the Bekenstein-Hawking entropy $S(r_{+})=\frac{A_{+}}{4G}$ as the maximal
entropy. This suggests that the range of values for the entropy of the
extremal  black hole in the (3+1)
dimensions is given by
Eq.~\eqref{Srange}  \cite{extremalshell}, as in the case of the 
rotating black hole in the (2+1) dimensions
considered here.

In Table I, we summarize the thermodynamic properties of the extremal
thin shell at its own gravitational radius with zero local limiting
temperature.

\begin{widetext}
\begin{table}[h]
\begin{minipage}[t]{1.0\textwidth}
\label{q4}
\begin{center}
\begin{tabular}{|c|c|c|c|c|c|c
}
\hline
    $T_0$ \,$b$\,\,
 &  $T$  \,$\beta$\,
 &  Backreaction 
 &  $V$
 &  Entropy 
\\ 
\hline
     $0$ \,$\infty$
  &  $0$ \,$\infty$
 &  Finite
 &  $1$ 
 &  $0\leq S(A_+)\leq \frac{A_+}{4G}$
\\
\hline
\end{tabular}
\end{center}
\caption{
The extremal shell with zero local temperature
at its own gravitational radius.}
\end{minipage}
\end{table}
\end{widetext}

So, we have found through a thin shell approach that
the (2+1)-dimensional extremal rotating BTZ black hole has an entropy 
$S=S(A_+)$ as we had found for the (3+1)-dimensional
 extremal Reissner-Nordstr\"om electrically charge black hole 
\cite{extremalshell,lqzn}, and again suggested 
$0\leq S(A_{+})\leq \frac{A_+}{4G}$, see Eq.~(\ref{Srange}).
The extremal black hole entropy was discussed originally
in a (3+1)-dimensional black hole context. 
It was found in \cite{ebh1} that the entropy of an extremal
(3+1)-dimensional black hole is zero, $S=0$, Eq.~(\ref{ent1ebh}).
This proposal
was substantiated by topology arguments.  The (2+1)-dimensional metric
does not contain one of the angle coordinates when compared to the
(3+1)-dimensional metric, but this is unessential since the main
arguments concern the topology of the $(
\tau,r)$ submanifold, where
$
\tau\equiv it$ is the Euclidean time, see also~\cite{ebh2} for
(2+1)-dimensional BTZ black hole.  
This reasoning is purely classical,
and inclusion of the backreaction due to quantum fields can destroy this
picture. On the other hand, the proposal put forward by string theory
leads to $S= \frac{A_+}{4G}$, Eq.~(\ref{ent1bh}), i.e.,
to a Bekenstein-Hawking entropy for extremal black
hole~\cite{string1,string2}, see~\cite{bss} for the BTZ black
hole. Our conclusion that 
$0\leq S(A_{+})\leq \frac{A_+}{4G}$, see Eq.~(\ref{Srange}),
incorporates both the $S=0$ and the 
$S= \frac{A_+}{4G}$ results.

\section{Entropy in the extremal BTZ black hole limit: 
Extremal thin shell with nonzero local temperature $T$
at the gravitational radius
} 
\label{sec5}

Now, we take again the extremal black hole limit but assume that the
extremal shell has another equation of state with a nonzero local
temperature. Specifically, at any $r>r_+$, we consider the following
temperature equation of state,
\begin{equation}
\label{T0bar}
T_0(r_+, R)={{\bar T_0}(r_+)}{k(r_+,R)}\,,
\end{equation}
where ${\bar T_0}$ is independent of $R$. 
Then, from Eq.~\eqref{temptolman0},
the temperature $T$ is 
$
T(r_+, R,r)={\bar T_0}(r_+)\frac{k(r_+,R)}{k(r_+,r)}
$.
At the shell $r=R$, one gets
\begin{equation}
\label{TnewR}
T(r_+, R)={\bar T_0}(r_+)\,.
\end{equation}
In terms of the inverse temperatures Eq.~(\ref{T0bar})
translates into
\begin{equation}
\label{bbar}
b(r_+, R)=\frac{\bar b(r_+)}{k(r_+,R)}\,,
\end{equation}
where
\begin{equation}\label{bbar2}
{\bar b(r_+)}\equiv\frac1{{\bar T_0}(r_+)}\,,
\end{equation}
is independent of $R$. 
Then, from Eq.~\eqref{temp_ext00},
the local inverse temperature $\beta$ at $r$ 
is 
$
\beta(r_+,R,r)={\bar b(r_+)}
\frac{k(r_+,r)}{k(r_+,R)}$.
Likewise, at
the shell, $r=R$, one  gets
\begin{equation}
\label{betabarR}
\beta(r_+,R)={\bar b(r_+)}
\end{equation} 
is constant and finite.

Combining \eqref{cdef} with \eqref{bbar},
we find that the rotational velocity equation of state 
can be written as  
\begin{eqnarray}
\label{barc}
1-g(r_+,R)=k(r_+, R)
 \left(1-{\bar g}(r_+)\right),
\end{eqnarray}
where we have defined 
\begin{eqnarray}
{\bar g(r_+)}
\equiv
1
-\frac{2G\ell^2}{r_+{\bar b} (r_+)} s(r_+),
\end{eqnarray}
which is independent of $R$ and is assumed to be
in the range $0<{\bar g} (r_+)<1$.

We now take the limit to the gravitational radius of the shell, $R\to
r_+$. In this limit, $k\to 0$. So, supposing $\bar T_0$ finite, which
we do, we see from Eq.~(\ref{T0bar}) that $T_0(r_+,R)=0$.  Since
$\bar T_0$ is finite, the local temperature $T$ at the shell is nonzero
but finite, since $T=\bar T_0$ 
from Eq.~\eqref{TnewR},
and hence the quantum backreaction remains finite even
when the shell is taken to its gravitational radius $R=r_+$.
Since $T$ is finite
$\beta$ given in Eq.~\eqref{betabarR}
is also finite, and
from Eq.~\eqref{sp}, we find that the shell with nonzero local
temperature rotates with thermal velocity less than the velocity of light $V<1$.
Finally, the
entropy of the shell obtained by integrating Eq.~\eqref{ent_ext2} is
well behaved and can take any function of $r_+$, or $A_+$.
Thus, using the same arguments as before, we can
write that the entropy of this extremal shell in the
extremal BTZ black hole limit also obeys
\begin{equation}  \label{Srange2}
0\leq S(A_{+})\leq \frac{A_+}{4G}\,.
\end{equation}

In Table II, we summarize the thermodynamic properties of the extremal thin
shell at its own gravitational radius with a local temperature $
T={\bar T_0}$. 
\begin{widetext}
\begin{table}[h]
\begin{minipage}[t]{1.0\textwidth}
\label{q42}
\begin{center}
\begin{tabular}{|c|c|c|c|c|c|c|
}
\hline
     $T_0$ \,$b$\,\,
 &  $T$ \,\,\,\,$\beta$\,
 &  Backreaction 
 &  $V$
 &  Entropy 
\\ 
\hline
    $0$ \,$\infty$
 &  Nonzero and Finite
 &  finite
 &  $<1$
 &  $0\leq S(A_+)\leq \frac{A_+}{4G}$
\\
\hline
\end{tabular}
\end{center}
\caption{
The extremal shell with nonzero local temperature at
its own gravitational radius.}
\end{minipage}
\end{table}
\end{widetext}


\section{Discussion on the angular, and the corresponding linear,
velocities of rotating thin shells}
\label{app1}

\subsection{Mechanical angular velocities}

It is instructive to rewrite the formula for the line element
outside of the shell (see Eqs.~(\ref{collect}) and (\ref{outq})) as
\begin{equation}  \label{collectout}
ds_{(o)}^2 =
-f_{(o)} (r)dt_{(o)}^2 
+ g_{(o)}(r) dr^2  
+r^2 \big(d\phi-\omega_{(o)} (r)dt_{(o)}\big)^2\,.
\end{equation}
Now, define
\begin{equation}
\label{omegaag}
\omega_R\equiv\omega_{(o)}(R)\,. 
\end{equation}
Static observers sitting at infinity, $r=\infty$, have an AdS metric,
since the BTZ metric turns into an asymptotically
AdS metric at infinity.  These
observers do not rotate relative to this AdS spacetime.  Thus,
observers sitting at infinity see a rotation of the shell with
$\omega_R$.  Here we keep the discussion quite general, for the
extremal BTZ case $\omega_R$  is given in Eq.~(\ref{shellrotout}), i.e.,
$\omega_R=\frac{r_+^2}{\ell R^2}$.

At the shell $r=R$, the metric (\ref{collectout})
becomes
\begin{equation}  \label{metshellagain}
ds_R^2 =-d\tau^2 +R^2d\psi^2, \quad R=R(\tau),
\end{equation}
where $\tau$ is the proper time at the shell 
and in terms of $dt_{(o)}$ is
$d\tau=\sqrt{f_{(o)} (R)}\,dt_{(o)}=k(r_+,R)\,dt_{(o)}$
(see also \eqref{reds}), 
and we have chosen 
a new angular coordinate $\psi$, 
\begin{equation}  \label{psiag}
\psi =\phi-\omega_R\, t_{(o)},
\end{equation}
such that the metric is displayed as 
diagonal (see also Eq.~(\ref{ct})).
The angular velocity $\omega_R$ defined in Eq.~(\ref{omegaag}),
appears thus quite 
naturally in Eq.~(\ref{psiag}) and is 
one of a number of interesting mechanical angular velocities
in this problem. Let us display the others, that we 
name $\bar \omega$, $\omega$, and 
$\omega_\infty$.

From Eqs.~(\ref{metshellagain})
and (\ref{psiag})
one deduces that
an observer comoving with the shell has $\psi={\rm constant}$.
Another observer on the shell moving with respect
to this comoving observer has angular velocity 
$\bar \omega$
with respect to the proper time on the shell $\tau$
given by
\begin{equation}  \label{angveltaupsi}
\bar\omega=\frac{d\psi}{d\tau}\,.
\end{equation}
This same observer has an angular velocity 
$\omega$ with respect to $t_{(o)}$ given by $\omega=\frac{d\psi}{dt_{(o)}}
=\frac{d\tau}{dt_{(o)}}\frac{d\psi}{d\tau}=k \,\bar\omega$,
where here $k\equiv k(r_+,R)$ to simplify the notation, i.e., 
\begin{equation}  \label{angveltpsi}
\omega=\frac{d\psi}{dt_{(o)}}=k\,\bar\omega\,. 
\end{equation}
Now, from Eq.~(\ref{collectout}), 
the coordinate $\phi$ is the angular coordinate 
defined at infinity.
Define then the angular velocity $\omega_\infty$
of an observer on the shell 
as seen by the coordinate $\phi$ 
and in terms of $t_{(o)}$ as
\begin{equation}  \label{angveltphi}
\omega_\infty=\frac{d\phi}{dt_{(o)}}\,. 
\end{equation}

We can now give a relation between 
$\omega$, $\omega_\infty$, and $\omega_R$, 
or between
$\bar\omega$, $\omega_\infty$, $\omega_R$,
and $k$.
Clearly, from Eq.~(\ref{psiag}),
\begin{equation}  \label{sumang}
\frac{d\psi}{dt_{(o)}} =\frac{d\phi}{dt_{(o)}}-\omega_R\,,
\end{equation}
so that, using Eqs.~(\ref{angveltpsi})
and (\ref{angveltphi})
in (\ref{sumang}),
we find
\begin{equation}  \label{sumang2}
\omega =\omega_\infty-\omega_R\,.
\end{equation}
If we prefer to use $\bar \omega$, i.e., 
to use the proper time coordinate
$\tau$, then 
from the help of Eqs.~(\ref{angveltaupsi})
and (\ref{angveltpsi}) we find
\begin{equation}  \label{sumang3}
\bar\omega =\frac{\omega_\infty-\omega_R}{k}\,.
\end{equation}
Expression~(\ref{sumang3}) is quite general. 
One can specialize. For instance, 
the special choice of $\bar \omega$ for which a shell's observer
detects no angular momentum flux density was given by Eq.~(39) of
\cite{energycond}, 
i.e., $\bar{\omega}=\frac{r_{-}}{r_{+}R}
\sqrt{\frac{R^{2}-r_{+}^{2}}{R^{2}-r_{-}^{2}
}}$. 
Note that $\omega_R$ is annotated as
$\Omega$  in \cite{energycond},
the angular velocity of the shell with respect to infinity.
We use here the notation $\Omega $ for a quite different quantity, the
thermodynamic angular velocity on the shell, see Eq.~\eqref{1st00}.
The linear velocities corresponding to $\bar\omega$, $\omega$, and
$\omega_\infty$ are $\bar v=\bar\omega R$,
$v=\omega R$, and  $v_\infty=\omega_\infty R$. Angular and linear
velocities 
share the same properties.

It is also instructive and important 
to point out the analogy with the black hole
case. In this 
context, Eq.~(\ref{sumang3}) is similar to the expression for the angular
velocity of the heat bath surrounding a
(2+1)-dimensional rotating black hole,
\begin{equation}  \label{omegaheatbathbh}
\omega_{\rm hb} =\frac{\omega_{\rm bh}-\omega_{\rm zamo}}{k}\,,
\end{equation}
where $\omega_{\rm hb}$ is the heat bath angular
velocity, $\omega_{\rm bh}$ is the black hole angular
velocity, $\omega_{\rm zamo}$
is the angular velocity of a zero angular momentum
observer (ZAMO), and $k$ in this formula is the redshift function
at the ZAMO radius, see Eq.~(13) of \cite{thermo_btz1}.
So, at a first glance, one may identify 
$\bar\omega\equiv\omega_{\rm hb}$,
$\omega_\infty\equiv\omega_{\rm bh}$, and 
$\omega_R\equiv\omega_{\rm zamo}$.
There is, however, an important difference
between Eqs.~(\ref{sumang3}) and (\ref{omegaheatbathbh}).  For a black
hole, the quantity 
$\omega_{\rm hb}$ enters both the mechanical and thermodynamic relations
(see \cite{thermo_btz1} for details), so the mechanical and
thermodynamic angular velocities coincide. However, for
a shell the relationship between both
angular velocities is much more subtle
as we now discuss.

\subsection{Mechanical and thermodynamic
angular and linear velocities for a rotating 
shell: The nonextremal and
extremal cases}
\label{sec6}

\subsubsection{The problem}

Consideration of thermal shells in  (2+1) dimensions
in the present paper as well in previous
ones \cite{energycond} (see also 
\cite{btzshell}) revealed interesting subtleties
otherwise hidden. At first sight, the shell thermodynamic
angular velocity $\Omega$ that appears
in the first law of thermodynamics, Eq.~\eqref{1st00},
should be immediately identified with the
shell 
mechanical angular velocity of a rotating shell fluid $\bar\omega$, Eq.~\eqref{sumang3},
in the same way as the quantity 
$\omega_{\rm hb}$ is both the mechanical and thermodynamic
angular velocity in the black hole case,
as discussed above (see \cite{thermo_btz1}).

However, the two angular velocities, namely, the thermodyamic
angular velocity $\Omega$ and the mechanical
velocity $\bar \omega$,
are indeed conceptually different, they have different
physical meanings. 
The quantity  $\Omega $  is the quantity ascribed to a
thermodynamic system as a whole, it is calculated from a pure thermodynamic
approach \cite{btzshell}. On the other hand, 
the quantity $\bar{\omega}$ represents the angular
velocity of the effective perfect fluid that fills the shell, i.e.,
it is obtained from geometry and mechanics, 
namely,  by
gluing two metrics, the BTZ metric and 
the zero mass BTZ metric on the different sides of the shell, calculating the
corresponding energy-momentum tensor and determining the angular velocity of
the effective fluid in terms of which 
the vanishing angular momentum flux is observed \cite{energycond}. 
The same rationale applies of course for
the corresponding linear velocities, $V=\Omega
R$ and $\bar v= \bar{\omega}R$, i.e., they are conceptually different.

So, a priori, it is not obvious at all, whether or not, and under what conditions,
these two velocities can be identified. It is thus instructive to compare this
issue for extremal and nonextremal shells.

\subsubsection{Extremal shell}
For an  extremal shell the situation is interesting
and different from what one would expect. 

Indeed, the two velocities $\Omega$ and 
$\bar \omega$, or $V$ and $\bar v$,
do not need to coincide
at all, as the
integrability condition (\ref{sp}) does not restrict the form of the
function $\Omega $ in the extremal case.
We will explain this fact now.

The thermodynamic angular velocity $\Omega$ presents new features. 
As it is argued above in Sec.~\ref{sec4}, when $T= 0$, then 
we must select
\begin{equation}
{\Omega}=\frac{1}{R} \,,\quad\quad\quad  T=\frac{1}{\beta }=0\,,
\label{omt2}
\end{equation}
so that,
\begin{equation}
V=1\,,
\label{veloc1}
\end{equation}
in the extremal limit under
discussion, to ensure the finiteness of
$s(r_{+})$. This is the extremal black hole limit
from an extremal thin shell with zero local temperature $T$,
see also Table I.
However, when $T\neq 0$, see Sec.~\ref{sec5},
any
thermodynamic angular velocity $\Omega$ obeying
\begin{equation}
{\Omega}<\frac{1}{R} \,,\quad\quad\quad  T=\frac{1}{\beta }\neq0\,,
\label{omt1}
\end{equation}
i.e., 
\begin{equation}
V<1\,,
\label{veloc<1}
\end{equation}
is suitable provided that $T=\beta ^{-1}$
remains nonzero in the extremal
black hole limit $R\rightarrow r_{+}$, see
Eq.~(\ref{sp}). This is the extremal black hole limit
from an extremal thin shell with nonzero local temperature $T$,
see also Table II.

The mechanical angular velocity $\bar\omega$, on the other hand, is given by
\begin{equation}
\bar\omega=\frac{1}{R}\,,  \label{ome}
\end{equation}
as  is seen from Eq.~(\ref{bar}). As a result, the linear mechanical
velocity, ${\bar v}=\bar\omega R$, is
\begin{equation}
\bar v=1\,,  \label{barv}
\end{equation}
which coincides with the velocity of light. This is quite
natural for a massless fluid, according to the discussion at the end
of Sec.~5 in \cite{energycond}.

\subsubsection{Nonextremal shell}

For a nonextremal shell, the situation is
what one expects upon imposing a reasonable regularity
condition. We will see this  now.

For nonextremal rotating shells,
from Eq.~(59) of \cite{btzshell} 
the thermodynamic angular velocity $\Omega$ 
is equal to
\begin{equation}
\Omega =\frac{r_{-}}{r_+R\sqrt{\left( 1-\frac{r_{+}^{2}}{R^{2}}\right)
\left( 1-\frac{r_{-}^{2}}{R^{2}}\right) }} \left[g(r_{+},r_{-})-\frac{r_+^2}{
R^{2}}\right],
\label{on}
\end{equation}
where again $g(r_{+},r_{-})$ is an arbitrary function of 
the gravitational and Cauchy radii, $r_{+}$ and $r_{-}$, respectively.
Now, if (i) the black hole limit $R\rightarrow r_{+}$ is reachable and,
(ii) we want to have $\Omega $ finite on the horizon, these conditions
select the choice $g=1$,
and hence 
\begin{equation}
\label{w}
\Omega=\frac{r_{-}}{r_{+}R}\sqrt{\frac{R^{2}-r_{+}^{2}}{R^{2}-r_{-}^{2}}}
\text{.}
\end{equation}
Then, $\Omega \sim \sqrt{R-r_{+}}\rightarrow 0$ in this limit.

The mechanical angular velocity $\bar{\omega}$
was obtained in Eq. (39) of \cite{energycond}
from matching conditions, namely,
\begin{equation}
\bar{\omega}=\frac{r_{-}}{r_{+}R}\sqrt{\frac{R^{2}-r_{+}^{2}}{R^{2}-r_{-}^{2}
}}\text{.}  
\label{bar}
\end{equation}
As we can see, in general, i.e., without imposing any
condition,
$\Omega$ given in Eq.~(\ref{on})
is different from 
$\bar\omega$ given in 
Eq.~(\ref{bar}),
$\Omega \neq \bar{\omega}$, so thermodynamic and mechanical 
angular velocities
do not coincide. 
The same applies for
the corresponding linear velocities, i.e., $v=\Omega
R$ and $\bar v= \bar{\omega}R$, do not coincide in general.
However, for the choice $g=1$, one gets  Eq.~(\ref{w})
that ensures the black hole
limit, and both Eq.~(\ref{w})
and Eq.~(\ref{bar}) coincide in this case, i.e., $\Omega =\bar{\omega}$.

To some extent, the situation resembles that with the
temperature. Then, $ T_{0}\neq T_{H}$ in general, where $T_{0}$ is the
temperature of the shell measured at $r_0$ in the AdS 
case and $T_{H}$ is the
Hawking temperature of a black hole with the same mass and other
parameters. However, if the
black hole limit is to be possible, we must take $T_{0}=T_{H} $ since otherwise
an infinite backreaction would destroy the horizon, see
\cite{charged} for more detailed discussion
in the asymptotically flat case. Thus, in both
cases, for the temperature and angular velocity, the 
existence of a well-defined black hole limit pushes forward the
regularity conditions that make the choice of these quantities
unambiguous.

\section{Conclusions}
\label{sec7}

We have investigated the thermodynamic entropy of an extremal rotating thin
shell in the (2+1)-dimensional asymptotically AdS spacetime, where the
exterior and interior of the shell were taken to be the
BTZ spacetimes and the ground state 
AdS spacetime, respectively. The matching procedure of these two
geometries is quite similar to the nonextremal shells, but nevertheless,
the thermodynamic properties of the
extremal shell have been shown
to be quite different from those of the nonextremal shell.

For the extremal rotating shells, the
thermodynamic
state independent variables, namely, the local proper mass $M$ of the shell, 
the position $R$ of the shell, and the angular momentum $J$ of the shell,
which appear
in the first law of thermodynamics, are not independent but are constrained by 
the relation $J=MR$.
Then, it was shown that the thermodynamic integrability
condition does not restrict the form of the inverse temperature $\beta$ and
rotational velocity $V$ equations of state of the shell, except
that the product $\frac{\beta}{R}(1-V)$ must be solely a function of $r_+$.
This also leads to the fact that the entropy $S$ is a function of
the gravitational radius $r_+$ alone, $S=S(r_+)$, or if one prefers, the
gravitational area $A_+$, $S=S(A_+)$. 
To find the temperature distribution throughout the spacetime
one must resort to the Tolman temperature formula. 

We have considered two specific classes of equations of state
for the temperature, one in which the local temperature
at the shell is zero and the other in which the local temperature
at the shell is nonzero and finite. We then took
appropriately the extremal black hole limit and found 
that in both cases the extremal black hole entropy is 
$S=S(A_+)$, arguing convincingly that one should
set $0\leq S(A_+)\leq \frac{A_+}{4G}$, i.e.,
the extremal black hole entropy has values in between
zero and the maximum entropy, the Bekenstein-Hawking entropy 
$\frac{A_+}{4G}$. Thus, rather than having
just two entropies for the extremal black holes, as previous
results have debated, namely zero and $\frac{A_+}{4G}$, we have
shown here that the extremal BTZ black hole
entropy may have a continuous range of entropies, limited
by precisely those two entropies.
Surely, the entropy that a particular black hole picks
must depend on past processes, notably on how it
was formed. In the gravitational collapse of a
shell to form an extremal black hole,
the entropy could depend on the equation
of state the shell, in the case of extremal black
hole pair creation, the entropy could depend
on the system initial conditions.

One can try to explain how
the entropy  of an extremal
black hole is ambiguous and can take a
range of values, $S=S(A_+)$.
The no hair theorems
for the the nonextremal BTZ black hole state that 
the final dynamical classical black hole is characterized
only by its horizon radius $r_+$ and Cauchy radius 
$r_-$ (or by the mass $m$ and angular momentum $\cal J$),
and for the extremal black hole 
only by $r_+$, as $r_+=r_-$
(or equivalently by the mass $m$, as ${\cal J}=m \ell$).
In the nonextremal case, the entropy is given in Eq.~(\ref{ent1bh}),
$S=\frac{A_+}{4G}$, and is a quantum, i.e., nonclassical,
quantity at its very heart, since 
the area, or perimeter in the (2+1) case, $A_+$
is measured in terms of Planck areas $1/G$ (recall the Planck constant
is put to unity).
Remarkably, this 
result is saying
that the black hole entropy, the variable that characterizes the macroscopic
quantum state, is characterized only by $A_+$
which itself is a function only of 
the variables of the no hair theorems, namely, $r_+$ and
$r_-$ (or $m$ and $\cal J$). 
For the extremal case the situation is different.
Although the area $A_+$ still appears in the entropy, the entropy
can be now a general function of the area itself, $S(A_+)$, and
does not need to be just proportional to it.
Thus, somehow the entropy for the extremal case
has more freedom, it can depend
on the initial object states that will form
the extremal black hole, as for instance on the equation
of state of the matter used to form the extremal black hole. 
At present, the origin of such an
unusual behavior in the extremal case, that the entropy depends on the
initial state of the system, is unknown, and a
detailed elucidation of this issue would be of significant physical
interest. 
Surely, fundamental theories to describe the entropy
have to take this into account.

Two further important and interesting features
have arisen in our analysis.
One feature that came about in analyzing the
shell with zero temperature is remarkable and should be mentioned. 
In this case  we have found a relation between the
limits $T\rightarrow 0$ and $V\rightarrow 1$, i.e., between the
the impossibility of reaching
absolute zero as the third law of
thermodynamics states, and the 
impossibility for a massive
body to reach the velocity of light.
Another interesting feature that 
our considerations revealed and that deserves
further analysis in a wider context of gravitational thermodynamics, not
only for the (2+1)-dimensional spacetimes,
is the relationship
between two angular, or between two linear, velocities, namely,
the mechanical velocity $\bar v$ and the thermodynamic velocity $V$.
For the extremal rotating shells that we have studied
this is especially pronounced
since $\bar v=1$, while the thermodynamic velocity $V$ can take any
value equal to or less than $1$.

\vskip 1cm

\appendix
\section{Tolman temperature and inverse
temperature formulas for a spherical shell: General discussion,
and the extremal and nonextremal cases}
\label{tolmangeneral}

\subsection{Tolman inverse temperature formula for a spherical
shell: The general formula}
The Tolman temperature formula
states that
the coordinate dependence of the temperature
in a static background with radial coordinate
$r$ is 
\begin{equation}  \label{temp_app0}
T=\frac{T_0}{k(r)},
\end{equation}
where $T_0$ depends not on $r$ but on some
possible parameters related to the system,
$T$ depends on $r$ through $k(r)$ only
and on these possible parameters
of the system,
and $k(r)$ is the redshift function that
depends essentially on $r$ 
but also can possibly
depend on the other possible parameters.
For the inverse temperatures $\beta$ and $b$,
with $\beta=1/T$, 
$b=1/T_0$, formula (\ref{temp_app0}) turns into
\begin{equation}  \label{invtemp_app0}
\beta=b\,k(r)\,.
\end{equation}
We will work with the inverse temperatures and the formula
Eq.~(\ref{invtemp_app0}). By inverting it,
we can always revert to the relation between
the temperatures themselves, Eq.~(\ref{temp_app0}).

We start with a general
nonextremal rotating BTZ spacetime in which a 
shell, or ring, is immersed at radius $R$. The discussion
is quite general and, with small modifications,
also holds for spacetimes with a spherical
shell
in other dimensions.
For such a spacetime, 
the gravitational radius $r_+$, the Cauchy radius $r_-$, and 
the radius of the shell $R$ are the parameters
that characterize the system.
Thus, in general, for a  nonextremal rotating BTZ spacetime
with a shell at radius $R$,
the Tolman general formula  Eq.~(\ref{invtemp_app0}) reads
in this case
\begin{equation}  \label{invtemp_app01}
\beta(r_+,r_-,R,r)=b(r_+,r_-,R)\,k(r_+,r_-,r)\,,
\end{equation}
where now $k(r_+,r_-,r)$ for the nonextremal shell
can be written explicitly as
\begin{eqnarray}  \label{knext0}
k\big(r_+,r_-,r\big) =\frac{r}{\ell} \sqrt{ \Big(1-\frac{r_+^2}{r^2}\Big) 
\Big(1-\frac{r_-^2}{r^2}\Big)}.
\end{eqnarray}
For a given
set of the parameters $(r_+,r_-,R)$
we can consider the Tolman inverse temperature formula
Eq.~(\ref{invtemp_app01})
which gives the inverse temperature $\beta$ at a coordinate $r$
knowing the inverse temperature $b$ at the radius where 
$k=1$. I.e.,  $\beta$ is the 
inverse temperature at $r$, blueshifted or redshifted 
with factor $k$ 
from the inverse temperature $b$ at the position where 
$k=1$.
One can then change the set 
$(r_+,r_-,R)$ and consider again 
the Tolman formula for the inverse temperature at $r$ for this new setting.

\subsection{Tolman inverse temperature formula for a spherical shell:
The extremal case}

One can now treat the ab initio extremal case.
For that, we take the extremal limit of
the nonextremal shell, $r_+\to r_-$. 
Then Eq.~\eqref{invtemp_app01}
gives in this limit
\begin{eqnarray}  \label{temp_ext0ext}
\beta(r_+,R,r)=b(r_+,R)k (r_+,r),
\end{eqnarray}
where now from Eq.~\eqref{knext0}
one has
$k\big(r_+,r\big) =\frac{r}{\ell}\Big(1-\frac{r_+^2}{r^2}\Big)$.
As before, the 
function $b$ can be interpreted as the inverse temperature of the
shell at the position $r_0$, where $k(r_+,r_0)=1$
(see also Eq.~\eqref{rastr}).
At the shell, $r=R$, the Tolman formula is
\begin{eqnarray}  \label{temp_ext0extR}
\beta(r_+,R)=b(r_+,R)k (r_+,R)\,,
\end{eqnarray}
where
\begin{eqnarray}
k\big(r_+,R\big) =\frac{R}{\ell}\Big(1-\frac{r_+^2}{R^2}\Big)\,.
\end{eqnarray}
There is no independent
thermodynamic integrability condition for 
$\beta$ or $b$ and so the general equation for the 
thermodynamic shell is $b=b(r_+,R)$.

One could expect that the dependence on $R$ could be
dropped, i.e., $b=b(r_+)$, which would be the case if 
one took directly
the limit to an extremal case from a nonextremal
case. But for an ab initio extremal shell, this does not
happen, and one has indeed $b=b(r_+,R)$.
This is the reason why extremal shells
are different thermodynamic systems
from nonextremal shells
and extremal black holes
are different thermodynamic systems from 
nonextremal black holes.
See also Eqs.~(\ref{temp_ext00}) and (\ref{temp_ext}) 
in the main text.

In the above discussion, we implied that $R>r_+$, the quantity
$b$ being finite. If a black hole limit is allowed, there are
restrictions that force us to take $b=\infty$, see Secs.~IV and V,
and Tables I and II.

\subsection{Tolman inverse temperature formula for a spherical shell:
The nonextremal case}

One can show that in the nonextremal case
Eq.~(\ref{invtemp_app01}) simplifies to
\begin{equation}  \label{invtemp_app01integrab}
\beta(r_+,r_-,R,r)=b(r_+,r_-)\,k(r_+,r_-,r)\,,
\end{equation}
i.e., the $R$ dependence in $b(r_+,r_-,R)$ drops,
and one gets instead simply $b(r_+,r_-)$, with
$k(r_+,r_-,r)$ still being given by
Eq.~(\ref{knext0}).

Indeed, at $r=R$, the Tolman formula~(\ref{invtemp_app01})
is $\beta(r_+,r_-,R,r=R)=b(r_+,r_-,R) k(r_+,r_-,r=R)$, i.e.,
$\beta(r_+,r_-,R)=b(r_+,r_-,R)\,k(r_+,r_-,R)$
with 
$k\big(r_+,r_-,R\big) =\frac{R}{\ell} \sqrt{ \Big(1-\frac{r_+^2}{R^2}\Big) 
\Big(1-\frac{r_-^2}{R^2}\Big)}$.
This is the Tolman inverse temperature
expression at the shell radius $R$.
If the shell happens to be at 
$R =R_0$, where 
$R_0$ is the radius at which
$k(r_+,r_-,R_0)=1$,
then at $R_0$,
$\beta(r_+,r_-,R_0)=b(r_+,r_-,R_0)$.

Now, it
is astonishing that the thermodynamic integrability conditions 
for the nonextremal BTZ shell case (this feature also holds
for shells in other dimensions)
give an equation which is a particular case of 
$\beta(r_+,r_-,R)=b(r_+,r_-,R)\,k(r_+,r_-,R)$. Indeed,
for the nonextremal rotating shells with two gravitational radii, $r_+$ and $
r_-$ ($r_+>r_-$), the integrability conditions obtained from the first law 
of thermodynamics yield that the local inverse
temperature equation of state of the shell \cite
{btzshell} is 
\begin{eqnarray}  \label{temp}
\beta(r_+,r_-,R)=b(r_+,r_-)\, k (r_+,r_-,R) ,
\end{eqnarray}
where $b(r_+,r_-)$ is an arbitrary function of $r_+$ and $r_-$ and
depends on the matter of the shell, and again $k$ is given by
\begin{eqnarray}  \label{knext0R}
k\big(r_+,r_-,R\big) =\frac{R}{\ell} \sqrt{ \Big(1-\frac{r_+^2}{R^2}\Big) 
\Big(1-\frac{r_-^2}{R^2}\Big)}\,,
\end{eqnarray}
\\
\noindent
which represents the gravitational redshift factor
in the nonextremal BTZ spacetime. Note that here, from the
integrability conditions, one finds $b=b(r_+,r_-)$, see
Eq.~(\ref{temp}), whereas directly from Tolman formula, one has
$b=b(r_+,r_-,R)$. Thus, thermodynamics,
through its integrability conditions, restricts the parameter space for
$b$. Equation~(\ref{temp}) is still the Tolman temperature
equation for the shell at the shell, but a restricted form.
Thus, by continuity, one also recovers the Tolman temperature
formula for the whole spacetime displayed in
Eq.~(\ref{invtemp_app01integrab})

Note again that the extremal limit of Eq.~(\ref{invtemp_app01integrab}),
i.e., of a nonextremal shell, is 
$\beta(r_+,R,r)=b(r_+)\,k(r_+,r)$, a particular,
restrictive, case of the general ab inito extremal shell 
Eq.~(\ref{temp_ext0ext}). For extremal shells, the correct expression
is, of course, Eq.~(\ref{temp_ext0ext}).

\centerline{}
\centerline{}
\centerline{}

\section*{ACKNOWLEDGEMENTS}

We thank Funda\c c\~ao para a Ci\^encia e Tecnologia
(FCT), Portugal, for financial support through
Grant~No.~UID/FIS/00099/2013.~JPSL thanks 
Coordena\c c\~ao de Aperfei\c coamento do Pessoal de
N\'\i vel Superior (CAPES),
Brazil, for support within the Programa CSF-PVE,
Grant No.~88887.068694/2014-00. MM~thanks~FCT, Portugal, 
for financial support through Grant No.~SFRH/BPD/88299/2012.
OBZ has been partially supported by the Kazan
Federal University through a state grant for scientific
activities.

\end{document}